\documentclass[twocolumn,superscriptaddress,amsmath,ammsymb,prb]{revtex4-2}
\usepackage{CJK}
\usepackage{epsfig}
\usepackage{amsmath}
\usepackage{amssymb}
\usepackage{graphicx}
\usepackage{dcolumn}
\usepackage{bm}
\usepackage[usenames]{color}
\usepackage[breaklinks]{hyperref}
\usepackage{soul}
\usepackage{ulem}
\hypersetup{colorlinks=true,allcolors=blue}

\begin{document}
	\begin{CJK*}{UTF8}{gbsn}
		
		\title{Universal scalefree non-Hermitian skin
			effect near the Bloch point}	
		\author {Wei Li ({\CJKfamily{gbsn}李维})}
		\author{Zhoujian Sun}
		\author{Ze Yang}
		\author{Fuxiang Li}
		\email[Corresponding author: ]{fuxiangli@hnu.edu.cn}
		\affiliation{School of Physics and Electronics, Hunan University, Changsha 410082, China}
		\date{\today}	
		\begin{abstract}
			The scalefree non-Hermitian skin effect (NHSE) refers to the phenomenon that the localization length of skin modes scales proportionally with system size in non-Hermitian systems. Authors of recent studies have demonstrated that the scalefree NHSE can be induced through various mechanisms, including the critical NHSE, local non-Hermiticity, and the boundary impurity effect. Nevertheless, these methods require careful modeling and precise parameter tuning. In contrast, in this paper, we suggest that the scalefree NHSE is a universal phenomenon, observable in extensive systems if these systems can be described by non-Bloch band theory and host Bloch points on the energy spectrum in the thermodynamic limit. Crucially, we discover that the geometry of the  generalized Brillouin zone determines the scaling rule of the localization length, which can scale either linearly or quadratically with the system size. In this paper, we enriches the phenomenon of the scalefree NHSE. 
		\end{abstract}
		\date{\today}	
		\maketitle	
	\end{CJK*}

	\section{\label{sec:level1}INTRODUCTION}
	In contrast to Hermitian operators, which guarantee real eigenvalues and conserve probability, non-Hermitian Hamiltonians provide a more general framework for the study of open or dissipative systems. This paradigm unveils previously unexplored physical phenomena that lack Hermitian analogs~\cite{edgeburst, 2, 3, 4, 5, 6, 7, 8, a1, a2, a4, pseudo} and has a wide range of applications including quantum sensing~\cite{sensor1, sensor2, sensor3}, topological photonics~\cite{pho1, pho2, pho3}, and signal amplification~\cite{sa1, sa2, sa3, sa4, sa5}. The most unique phenomenon observed in non-Hermitian systems is the non-Hermitian skin effect (NHSE) ~\cite{nhse1, nhse2, nhse3, nhse4, nhsex, nhsexx, nhse5, nhse6, obc1, obc2, obc3, obc4, obc5, obc6, obc7, obc8, BP}, in which the eigenstates of the system are exponentially localized at the boundaries under open boundary conditions (OBCs). To comprehensively account for the NHSE and the corresponding energy spectrum under OBCs, a complete theory, known as non-Bloch band theory~\cite{band, nhse5, aGBZ}, has been established.
	
	Recent theoretical and experimental work has expanded our understanding of the NHSE. A characteristic of the NHSE has been identified, wherein the localization length of skin modes scales with the system size, a phenomenon termed the scalefree NHSE or scalefree localization. This effect has been observed in systems that exhibit the critical NHSE~\cite{sfl1, sfl2}, local non-Hermiticity~\cite{sfl3, sfl4}, or boundary impurity effect~\cite{sfl5}. Nevertheless, such systems require careful modeling and fine parameter tuning, manifesting remarkable alterations in the energy spectrum configuration as the system size varies.
	
	In this paper, we uncover that the scalefree NHSE is a ubiquitous phenomenon, observable in extensive systems if these systems can be described by non-Bloch band theory and host Bloch points on the energy spectrum in the thermodynamic limit. The appearance of Bloch points stems from the intersection of the generalized Brillouin zone (GBZ) with the conventional Brillouin zone (BZ), which can be achieved by adjusting the parameters of the system. Within these usual models, the energy spectrum configuration remains largely invariant with respect to system size, thereby allowing us to elaborate the scaling rule of this type of scalefree NHSE via non-Bloch band theory. To emphasize the universality of the scalefree NHSE, we derive formulas that accurately describe the power-law behaviors near the Bloch point and verify the applicability of these formulas in more complicated models. Crucially, based on these formulas, we demonstrate that different geometries of the GBZ yield scalefree power-law behaviors of inverse localization length $\kappa_m$ with increasing system size $L$: $\kappa_m \sim L^{-j}$ with distinctive exponents $j$ that can be not only $1$ but also $2$. In this paper, we provide insight into the physical mechanisms behind the scalefree NHSE.
	
	\section{GBZ and Bloch point}
	
	Without loss of generality, we begin with a single-band tight-binding non-Hermitian model in one dimension with the Bloch Hamiltonian:
	\begin{equation}
		H(k) =t_{-2}e^{-2ik}+t_{-1}e^{-ik}+t_1e^{ik},
		\label{eq1}
	\end{equation}
	where $k$ denotes the Bloch wave number that resides in the BZ, with $k \in [-\pi, \pi]$. For simplicity, the hopping parameters $t_{-2}$, $t_{-1}$, and $t_1$ are set to be real. The Bloch Hamiltonian describes non-Hermitian systems with periodic boundary conditions (PBCs), but it fails to represent non-Hermitian systems with OBCs due to significant spectral differences caused by the presence of skin modes. To elucidate the nature of these non-Hermitian skin modes, the concept of the generalized Bloch phase factor $\beta \equiv e^{ik-\kappa(k)}$~\cite{nhse5} has been introduced,  where the NHSE inverse localization length $\kappa(k)=-\rm ln|\beta|$ is a real number representing the degree and direction of the skin modes. In the thermodynamic limit, the OBC spectrum $E_{\rm OBC}$ is determined by evaluating the generalized Bloch Hamiltonian $H(\beta)$ along a uniquely defined contour, termed the GBZ. The $H(\beta)$ corresponding to Eq.~(\ref{eq1}) is given by
	\begin{equation}
		H(\beta)=t_{-2}\beta^{-2}+t_{-1}\beta^{-1}+t_1\beta.
		\label{eq2}
	\end{equation}
	The OBC spectrum $E_{\rm OBC}$ and the GBZ of $H(\beta)$ can be determined by solving the following characteristic equation:
	\begin{equation}
		f(\beta,E_{\rm OBC}) =H(\beta)-E_{\rm OBC}=0.
		\label{eq3}
	\end{equation}
	For this model, the characteristic equation yields three solutions $\beta_{j}$ with $j=1, 2, 3$ for a given $E_{\rm OBC}$. The GBZ can be determined by setting $|\beta_2|=|\beta_3|$ after one orders the three solutions in ascending magnitude as $|\beta_1|\le|\beta_2|\le|\beta_3|$ (see Appendix~\ref{sec:aA} for detailed analysis). In Figs.~\ref{fig1}(a) and \ref{fig1}(b), the GBZ and the OBC spectrum $E_{\rm OBC}$ are plotted as colored lines, whereas the BZ and the PBC spectrum $E_{\rm PBC}$ are plotted as dashed lines. As evident from Fig.~\ref{fig1}(a), the GBZ intersects the BZ at two distinct points, one of which is denoted as $\beta_B$ for later use. These intersection points correspond to an identical energy level, referred to as the Bloch point $E_B$. As elaborated in Ref.~\cite{BP}, $E_B$ is the intersection of the spectra $E_{\rm PBC}$ and $E_{\rm OBC}$, as shown in Fig.~\ref{fig1}(b). 
	
	\begin{figure}[tb]
		\centering
		\includegraphics[width=0.48\textwidth]{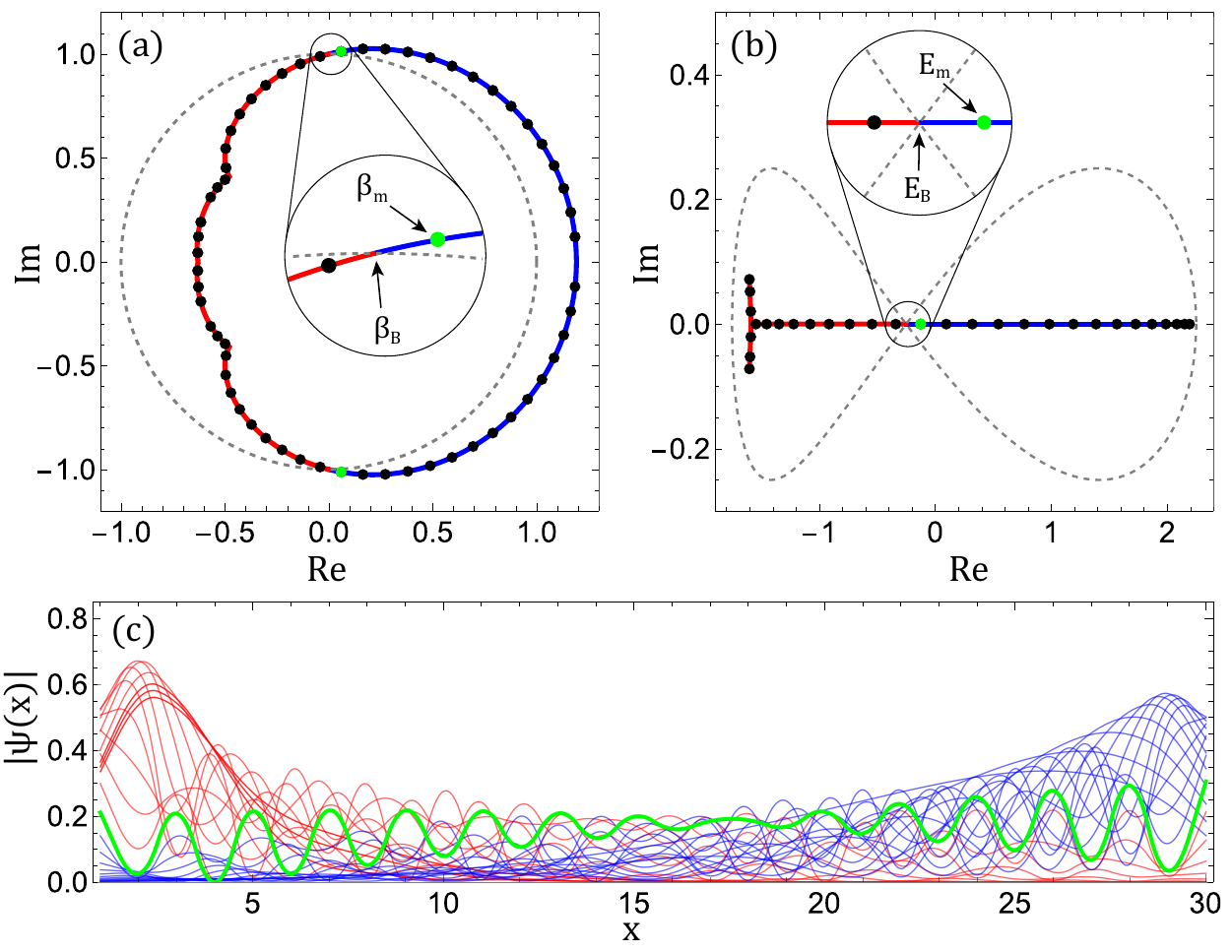}\\
		\caption{(a) Generalized Brillouin zone (GBZ; solid curve) and Brillouin zone (BZ; dashed unit circle) are plotted for the model in Eq.~(\ref{eq1}). The part of the GBZ inside (outside) the BZ is colored red (blue). (b) Open-boundary-condition (OBC) spectrum $E_{\rm OBC}$ (colored curve) and periodic-boundary-condition (PBC) spectrum $E_{\rm PBC}$ (dashed loop). The generalized Bloch phase factors and eigenenergies of the open chain with length $L=30$ are plotted as dots in (a) and (b), where the green dots correspond to the specific energy level $E_m$ that we are interested in. (c) Eigenfunctions corresponding to the dots on the red (blue) part of the GBZ are plotted as red (blue) curves, where the green curve corresponds to the energy level $E_m$. In all plots, parameters are set as $t_{-2}=1/4$, $t_{-1}=1$, and $t_1=1$.}\label{fig1}
	\end{figure}
	
	In addition to determining the OBC spectrum $E_{\rm OBC}$, the GBZ also provides valuable information about the eigenstates under OBCs. Typically, the GBZ is not a unit circle like the BZ but has some irregular structure. The segment of the GBZ that lies inside (outside) the unit circle corresponds to a set of the left (right)-localized skin modes~\cite{nhse5,nhse6}. At the energy level $E_B$, the eigenstate extends as a Bloch wave owing to $|\beta_B|=1$ and $\kappa = 0$. Away from the Bloch point, the eigenstates are all localized either on the right or left end of the system under OBCs. In the model in Eq.~(\ref{eq2}) for a system of size $L=30$, the generalized Bloch phase factors $\beta$ and eigenenergies $E$ are denoted by discrete dots in Figs.~\ref{fig1}(a) and \ref{fig1}(b), respectively. Moreover, $|\Psi(x)|$ the normalized wave-function values of all eigenstates at $x$ are plotted in Fig.~\ref{fig1}(c), where the left (right)-localized eigenstates are colored red (blue).
	
	\begin{figure}[tb]
		\centering
		\includegraphics[width=0.48\textwidth]{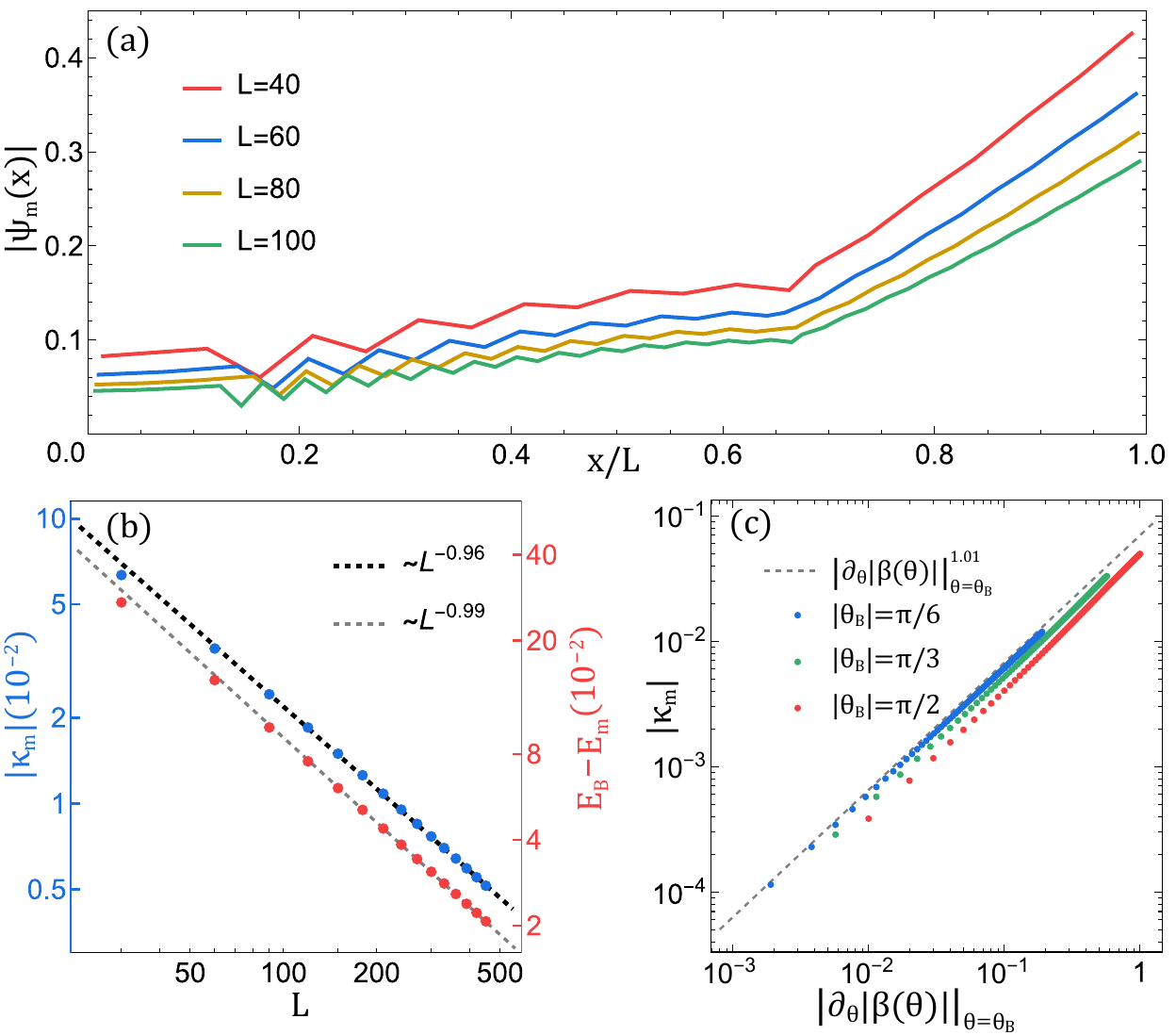}\\
		\caption{(a) The rescaled profiles of the wave function $\Psi_m(x)$ at energy level $E_m$ with system sizes $L=40$, $60$, $80$, and $100$ (red, blue, orange, and green, respectively) are displayed. The peaks of the original wave functions are extracted and depicted for clarity. (b) The inverse localization length $\kappa_m $ (left ticks, blue) and the difference between $E_m$ and $E_B$ (right ticks, red) are shown to decrease as $\sim L^{-1}$. Parameters are $t_{-2}=t_{-1}=t_1=1$ for both (a) and (b). (c) The dependence of $|\kappa_m|$ on $\big|\partial_\theta| \beta(\theta)|\big|_{\theta=\theta_B}$ is plotted. The red, green, and blue dots correspond to three cases where $| \theta_B|=\pi/2$, $\pi/3$, and $\pi/6$, respectively, and the system size is $L=40$. }\label{fig2}
	\end{figure}
	
	\section{Scalefree NHSE in single-band systems}
	
	\subsection{Scalefree NHSE near Bloch point}
	We proceed to investigate the occurrence of the scalefree NHSE near the Bloch point. After arranging the energy levels based on the real parts of the eigenenergies, we focus our attention on the lowest energy level above the Bloch point, labeled $E_m$, for subsequent analysis. (Later, we will show that the other energy levels nearest to the Bloch point in the complex plane also exhibit the scalefree NHSE.) In Fig.~\ref{fig1}, the energy level $E_m$, its associated generalized Bloch phase factor $\beta_m$, and the magnitude of the wave function $\Psi_m(x)$ at $E_m$ are highlighted in green. Notably, $\beta_m$ is situated near the intersection between the GBZ and the conventional BZ, which leads to $\kappa_m$ being ~$0$, and as a result, the wave function $\Psi_m(x)$ is less localized in comparison with the other wave functions.
	
	Throughout our investigation, we calculated the eigenstate at the energy level $E_m$ for various system sizes. Remarkably, these eigenstates maintain similar configurations if the system size is taken as the measure of length, as illustrated in Fig.~\ref{fig2}(a). This suggests that, unlike usual exponentially decaying wave functions with fixed localization length, the localization length of the wave function at the energy level $E_m$ is proportional to the system size, a salient characteristic of the scalefree NHSE. It should be noted that variations in the amplitude of the wave functions across different system sizes are a consequence of wave function normalization. To further elucidate the scalefree power-law behavior, we calculate the inverse localization length $\kappa_m$ at energy level $E_m$ for different system sizes. As shown by the blue dots in Fig.~\ref{fig2}(b), the inverse localization length $\kappa_m$ is inversely proportional to the system size $L$, i.e., $\kappa_m\sim L^{-1}$. Such a unique scalefree eigenmode arises due to the convergence of $E_m$ and $E_B$ as the system size $L$ increases~\cite{sfl1, sfl2}. Specifically, $|E_B-E_m|\sim L^{-1}$, as denoted by the red dots in Fig.~\ref{fig2}(b).
	
	\subsection{Influence of the geometry of GBZ}
	As demonstrated in Ref.~\cite{PT}, the presence of cusps on the GBZ gives rise to non-Bloch van Hove singularities, thereby affecting the asymptotic behavior in the vicinity of the cusps. Given this, we conjecture that the geometry of the GBZ may also affect the power-law relation near the Bloch point. Our study is then extended to investigate the influence of the GBZ geometry on the scalefree NHSE. For this purpose, we first introduce $\beta_B=e^{i\theta_B}$ at the Bloch point and rewrite Eq.~(\ref{eq2}) as 
	\begin{equation}
		H\left(\theta_B \right) = t_{-2}e^{-2i\theta_B}+t_{-1}e^{-i\theta_B}+t_1e^{i\theta_B} =E_B.
	\end{equation}
	To simplify, we judiciously choose parameters to ensure that $E_B$ resides on the real axis. Under these conditions, we obtain the following expressions for $\theta_B$ and $E_B$:
	\begin{equation}
		\begin{split}
			\theta_B & = \pm\arccos\frac{t_1-t_{-1}}{2t_{-2}} ,\quad
			E_B  = \frac{t_1^2-t_{-2}^2-t_1t_{-1}}{t_{-2}},\\
		\end{split}
		\label{eq6}
	\end{equation}
	where we assume $\theta_B\in[-\pi/2, \pi/2]$. After parametrizing the GBZ as $\beta=|\beta(\theta)|e^{i\theta}$, the geometry of the GBZ near the Bloch point can be characterized by the magnitude of $\big|\partial_\theta| \beta(\theta)|\big|_{\theta=\theta_B}$. According to Eq.~(\ref{eq6}), we can modify the shape of the GBZ by varying the value of $\big|\partial_\theta| \beta(\theta)|\big|_{\theta=\theta_B}$ while keeping $\theta_B$ constant. Figure~\ref{fig2}(c) illustrates that $|\kappa_m|$ is proportional to $\big|\partial_\theta| \beta(\theta)|\big|_{\theta=\theta_B}$ when $L$ and $\theta_B$ are kept constant. The datasets, differentiated by color, indicate that the intersection locations between the GBZ and the BZ exert negligible influence on $|\kappa_m|$. Thus far, we have discovered that the geometry of the GBZ significantly impacts the scalefree NHSE. In Sec.~\ref{sec:s}, we will further examine this effect by analyzing three specific scenarios.
	
	\begin{figure*}[ht]
		\centering
		\includegraphics[width=0.8\textwidth]{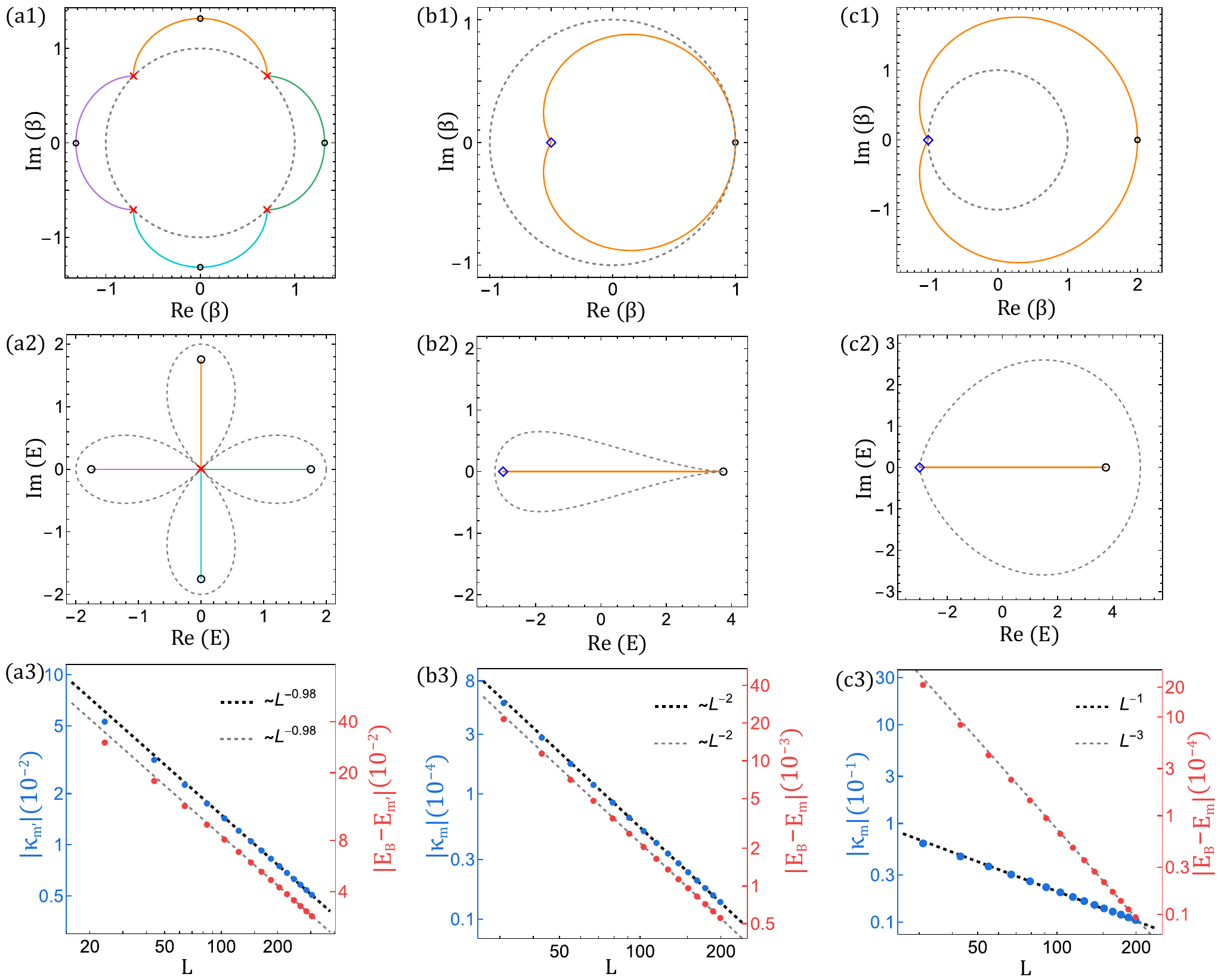}\\
		\caption{Generalized Brillouin zone (GBZ) and Brillouin zone (BZ), $E_{\rm OBC}$ and $E_{\rm PBC}$, and the scalefree power-law behaviors of inverse localization length and eigenenergy near the Bloch point are plotted for (a1)-(a3) $H\left(\beta\right)=\beta^{-3} +\beta^1$, (b1)-(b3) $H\left(\beta\right)=\beta^{-2}/4 +3\beta^{-1}/2+2\beta$, and (c1)-(c3) $H\left(\beta\right)=\beta^{-2} +3\beta^{-1}+\beta$. The parts of the GBZ and $E_{\rm OBC}$ that match in color correspond to each other. (a3), (b3), and (c3) exhibit three distinct scalefree power-law behaviors.}\label{fig3}
	\end{figure*}
	
	\subsection{Special cases of the geometry of GBZ\label{sec:s}}
	When modifying the geometry of the GBZ, three notable scenarios may arise: the BZ intersects the GBZ at a cusp, at a saddle point (satisfying $\partial_\beta H(\beta) = 0$~\cite{nhse1}), or at a higher-order saddle point that is also a cusp.
	
	In Fig.~\ref{fig3}(a1), the BZ passes right through the cusps of the GBZ, marked by the red crosses. These cusps correspond to discontinuous points of $\partial_\theta| \beta(\theta)|_{\theta=\theta_B}$. The GBZ is partitioned into different colored sections by these cusps, each corresponding to different branches of the OBC spectrum $E_{\rm OBC}$, as illustrated in Figs.~\ref{fig3}(a1) and \ref{fig3}(a2). Additionally, the branch point of the OBC spectrum marked by the red cross in Fig.~\ref{fig3}(a2) is also a Bloch point that corresponds to the four cusps of the GBZ. For our study, we select the energy level nearest to the Bloch point on the orange branch of the OBC spectrum and label it as $E_{m'}$. As the system size increases, the energy level $E_{m'}$ gradually approaches the Bloch point. Simultaneously, the generalized Bloch phase factor $\beta_{m'}$ at the energy level $E_{m'}$ converges toward the cusp and remains within the orange segment of the GBZ. Consequently, the discontinuity of $\partial_\theta| \beta(\theta)|_{\beta=\beta_B}$ at the cusps do not affect the scalefree NHSE. Figure~\ref{fig3}(a3) demonstrates the scalefree power-law behaviors of $|\kappa_{m'}|$ and $|E_B-E_{m'}|$. 
	
	In Fig.~\ref{fig3}(b1), the GBZ and the BZ are tangent at the saddle point of the GBZ, marked by a black circle. The Bloch point in this case is located precisely at the right endpoint of the OBC spectrum~\cite{aGBZ, PT}, as depicted by a black circle in Fig.~\ref{fig3}(b2). Here, we still use $E_m$ to represent the energy level closest to the Bloch point. It is worth noting that the scalefree power-law behaviors change significantly due to $\big|\partial_\theta| \beta(\theta)|\big|_{\theta=\theta_B}=0$ in this case. As illustrated in Fig.~\ref{fig3}(b3), both $|\kappa_{m}| $ and $|E_B-E_{m}|$ decrease as $\sim L^{-2}$, which is in stark contrast to the previous cases.
	
	In Fig.~\ref{fig3}(c1), an special intersection emerges between the GBZ and the BZ. Although it may appear as a cusp, it is, in fact, a third-order saddle point of the GBZ, denoted by a blue diamond. The $l$th order saddle point of the GBZ is defined as $H(\beta) -E_S=\partial_\beta H(\beta)\big|_{\beta=\beta_S}=\ldots =\partial_\beta^{l-1} H(\beta) \big|_{\beta=\beta_S} = 0$, where $\beta_S$ denotes the saddle point and $E_S$ denotes the saddle point energy. This unique geometry of the GBZ leads to the interesting phenomenon that the scalefree power-law behaviors of $|\kappa_m|$ and $|E_B-E_m|$ with respect to system size $L$ are no longer the same, as shown in Fig.~\ref{fig3}(c3).
	
	Upon analyzing these three cases, it becomes evident that the geometry of the GBZ significantly impacts the scalefree power-law behaviors of $|\kappa_m|$ and $|E_B-E_m|$. Intriguingly, we also observe that the scalefree power-law exponent of $|E_B-E_m|$ equals the order $l$ of the saddle point, i.e., $|E_B-E_m|\sim L^{-l}$. This finding motivates us to formulate equations for computing these exponents.
	
	\subsection{Derivation of scalefree power-law exponents \label{sec:A}}
	To obtain the exponents that govern the scalefree power-law behavior near the Bloch point, we now derive algebraic expressions. The density of states near the Bloch point can be determined as
	\begin{equation}
		\rho\left(E\right) \sim \left|E-E_B\right|^{-\alpha}.
		\label{pe}
	\end{equation}
	In the parity-time ($\mathcal{PT}$) symmetric phase, where $E_{\rm OBC}$ is purely real, the exponent $\alpha$ for an $l$th order saddle point is $\alpha=1-1/l$~\cite{PT}. Moreover, the eigenenergy $E_m$ can be expanded near the $\beta_B$ as
	\begin{equation}
		E_m\approx H(\beta_B)+ \left. \frac{\partial^{(l)}_\beta H(\beta)}{l!} \right| _{\beta=\beta_B} (\beta_m-\beta_B)^l.
		\label{em}
	\end{equation}
	This derivation yields the relationship between the eigenenergy $E_m$ and the inverse localization length $\kappa_m$, given by $E_m-E_B\sim (\beta_m-\beta_B)^l={\rm {exp}} (i\theta_Bl) \{{\rm {exp}} [i(\theta_m-\theta_B)-\kappa_m]-1\}^l$, where $\kappa_B$ is omitted since it is equal to $0$. To further eliminate the variables $\theta_m$ and $\theta_B$, we can express them in terms of $\kappa_{m}$. Employing the method utilized to define the saddle point order on the GBZ, we can set an order $j$ for $\kappa(\theta)$ such that $\partial_\theta \kappa(\theta)|_{\theta=\theta_B}=\ldots = \partial_\theta^{j-1}\kappa(\theta)|_{\theta=\theta_B} = 0$. It should be noted that the inverse localization length $\kappa(\theta) = -{\rm ln}|\beta(\theta)|$ can be obtained directly from the parametric equations of the GBZ. We then expanded $\kappa_m$ around $\theta_B$ to the $j$th order as $\kappa_m \approx ( \theta_m-\theta_B )^j/A$, where $A=[j!/\partial^{(j)}_\theta \kappa(\theta)] _{\theta=\theta_B}$. Utilizing these expressions, we establish that $|E_m-E_B|\sim |(iA\kappa_m^{1/j}-\kappa_m)^l|$, where we have used the equivalent infinitesimal substitution ${\rm {exp}}(iA\kappa_m^{1/j}-\kappa_m)-1\sim iA\kappa_m^{1/j}-\kappa_m$ since $\kappa_m\rightarrow0$ when system size $L\rightarrow\infty$. We note that $\kappa_m$ is of the same order or a higher-order infinitesimal than $\kappa_m^{1/j}$ since $j$ is a positive integer. This means that the relationship between the eigenenergy $E_m$ and the inverse localization length $\kappa_m$ can be rewritten as
	\begin{equation}
		|E_m-E_B|\sim|\kappa_m|^{l/j},
		\label{emkm}
	\end{equation}
	regardless of the value of the order $j$. Upon integrating both sides of Eq.~(\ref{pe}) over the energy interval $[E_B^+,E_m^+]$ and considering that only a single energy level (i.e., $E_m$) falls within this interval, the system size dependence of the energy difference $|E_m-E_B|$ can be described as:
	\begin{equation}
		|E_m-E_B|\sim L^{-l}.
		\label{eml}
	\end{equation}
	Now by substituting Eq.~(\ref{emkm}) into Eq.~(\ref{eml}), we deduce a scalefree power-law relationship between the inverse localization length $\kappa_m$ and the system size $L$:
	\begin{equation}
		|\kappa_m| \sim L^{-j}.
		\label{kml}
	\end{equation}
	Equations~(\ref{eml}) and~(\ref{kml}) accurately describe the power-law behaviors observed in Figs.~\ref{fig3}(a3), \ref{fig3}(b3) and \ref{fig3}(c3), with the exponents $l$ and $j$ determined by the geometry of the GBZ. Notably, although initially derived within the $\mathcal{PT}$-symmetric phase, our numerical findings affirm the applicability of these equations to the $\mathcal{PT}$-symmetry broken phase as well.
	
	\begin{figure}[t]
		\centering
		\includegraphics[width=0.48\textwidth]{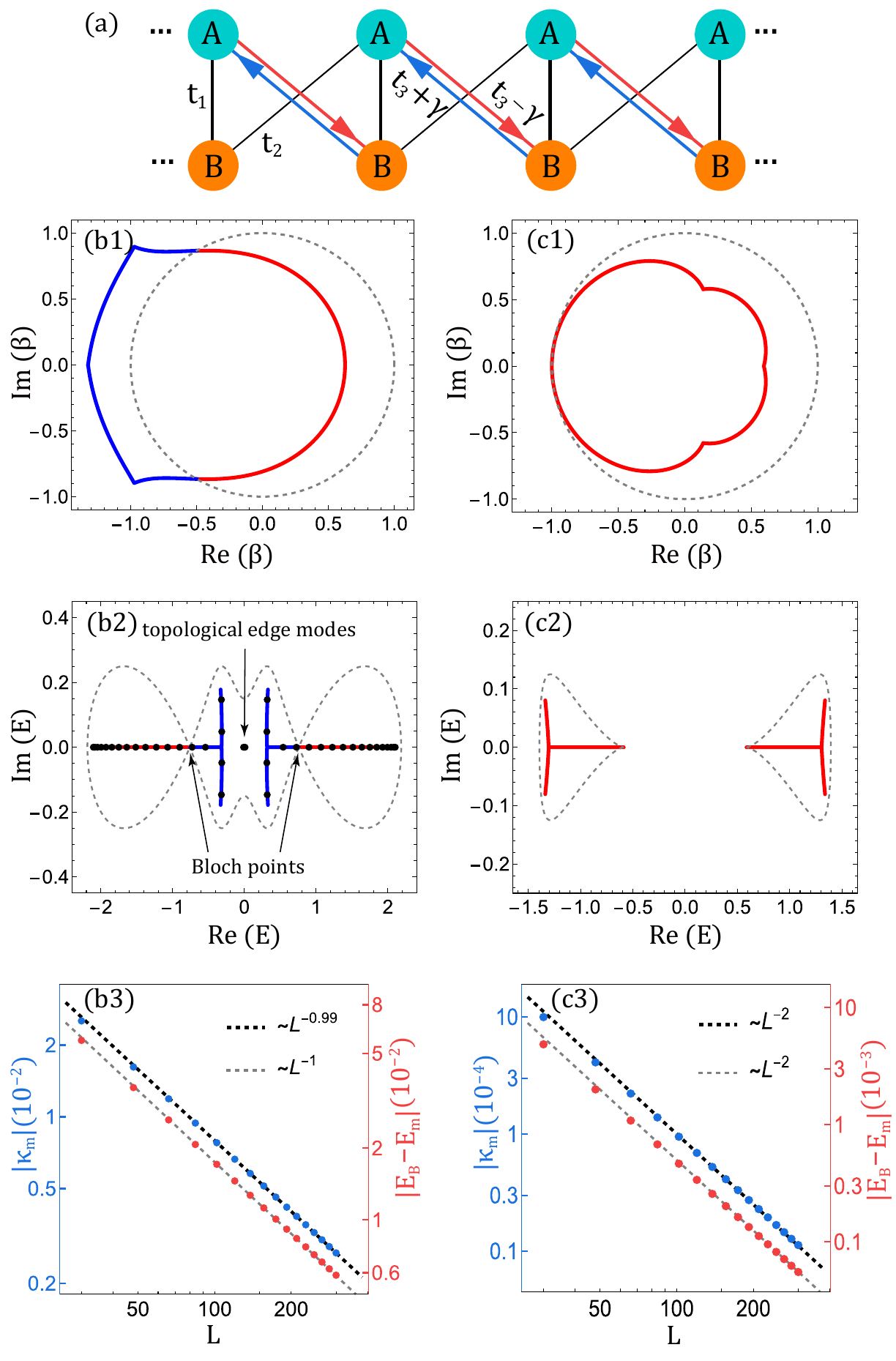}\\
		\caption{(a) Su-Schrieffer-Heeger (SSH) model with alternating $t_1$, $t_2$ hoppings and a non-Hermitian $t_3\pm \gamma$ hopping. (b1)--(b3) are plotted with parameters $t1 = t2 = 1$, $t3 = \frac{1}{5}$, $\gamma = \frac{1}{4}$. Black dots in (b2) represent the eigenenergies of the system with $L = 20$. (c1)--(c3) are plotted with parameters $t1 = 1$, $t2 = \frac{1}{2}$, $t3 = -\frac{1}{10}$, $\gamma = \frac{1}{4}$. These two different scalefree power-law behaviors shown by the two-band model are consistent with those in the single-band model.}\label{fig4}
	\end{figure}
	
	\section{Scalefree NHSE in multiband systems}
	
	Such a scalefree NHSE as we found near the Bloch point can also occur in multiband systems including the non-Hermitian Su-Schrieffer-Heeger (SSH) model, which is pictorially shown in Fig.~\ref{fig4}(a). The generalized Bloch Hamiltonian reads $H(\beta)=R_+(\beta)\sigma_++R_-(\beta)\sigma_-$, where $\sigma_\pm=(\sigma_x\pm i\sigma_y)/2$, and $R_\pm(\beta)$ are given by
	\begin{equation}
		\begin{split}
			R_+(\beta)&= t_1+t_2\beta^{-1}+(t_3+\gamma)\beta, \\
			R_-(\beta)&= t_1+t_2\beta+(t_3-\gamma)\beta^{-1}.
		\end{split}
		\label{eq13}
	\end{equation}
	Therefore, the eigenvalue equation can be written as
	\begin{equation}
		R_+(\beta)\ R_-(\beta)=E^2,
		\label{eq14}
	\end{equation}
	which is a quartic equation for $\beta$. Upon arranging the four solutions in ascending magnitude, denoted as $|\beta_1|\le|\beta_2|\le|\beta_3|\le|\beta_4|$, the GBZ is determined by the trajectory of $\beta_2$ and $\beta_3$ satisfying the condition $|\beta_2|=|\beta_3|$, as shown in Figs.~\ref{fig4}(b1) and \ref{fig4}(c1). The scalefree power-law exponents $l$ and $j$ assume the value of $1$ [Fig.~\ref{fig4}(b3)] when the GBZ merely intersects the BZ [Fig.~\ref{fig4}(b1)], and $2$ [Fig.~\ref{fig4}(c3)] when the GBZ is tangent to the BZ [Fig.~\ref{fig4}(c1)]. Interestingly, the Hamiltonian for this model manifests sublattice symmetry $\sigma_z^{-1}H(\beta)\sigma_z=-H(\beta)$, thereby ensuring the existence of topological edge modes under specific parameter conditions [Fig.~\ref{fig4}(b2)]. Nevertheless, the presence of zero-energy topological edge modes does not affect the scalefree NHSE, as the zero-energy mode resides within the bulk band gap. These numerical results are consistent with the theoretical predictions mentioned earlier.
	
	\section{Summary}
	In this paper, our study focuses on a type of scalefree NHSE that occurs near the Bloch point. We observe that the wave function at the energy level closest to the Bloch point exhibits a scalefree property as $L$ varies. This type of scalefree NHSE is a ubiquitous phenomenon, observable in extensive systems if these systems can be described by non-Bloch band theory and host Bloch points on the energy spectrum in the thermodynamic limit. Moreover, we find that the geometry of the GBZ affects the scalefree power-law behaviors near the Bloch point. We have derived formulas describing these scalefree power-law behaviors based on the Bloch energy band theory, thereby underscoring the universality of the scalefree NHSE. In contrast to the methods used in Refs.~\cite{sfl1, sfl2, sfl3, sfl4, sfl5}, which require careful modeling and fine parameter tuning, our approach only necessitates rough parameter adjustments to achieve the scalefree NHSE. We expect that our theoretical predictions can be verified in various non-Hermitian platforms, including electric circuit~\cite{sfl1, sfl5, ec1, ec2} and photonic crystals~\cite{pc1, pc2}, where the NHSE has been realized.
	
	\section*{Acknowledgements}
	This paper was supported by the National Key Research and Development Program of the Ministry of Science and Technology (Grant No. 2021YFA1200700), the National Natural Science Foundation of China (Grants No. 11905054 and No. 12275075), and the Fundamental Research Funds for the Central Universities of China.
	
	\appendix

	\section{Derivation of the GBZ form Eq.~(\ref{eq2}) \label{sec:aA}}
	In this appendix, we briefly explain the derivation of the GBZ of the single-band model in Eq.~(\ref{eq2}). In real space, the Hamiltonian of this model with the system size $L$ reads
	\begin{equation}
		\hat{H}= \sum_{i=1}^{L} t_{-2}\hat{c}_{i+2}^{\dagger}\hat{c}_{i} + t_{-1}\hat{c}_{i+1}^{\dagger}\hat{c}_{i} + t_{1}\hat{c}_{i}^{\dagger}\hat{c}_{i+1},
	\end{equation}
	and the real-space eigenequation leads to 
	\begin{equation}
		t_{-2}\psi_{n-2} + t_{-1}\psi_{n-1} + t_{1}\psi_{n+1} = E_{\rm OBC}\psi_{n},
	\end{equation}
	in the bulk of chain, where $n=3,4,\ldots,L-1$. The boundary conditions satisfy
	\begin{equation}
		\begin{split}
			t_{1}\psi_{2} &= E_{\rm OBC}\psi_{1},\\
			t_{-1}\psi_{1} + t_{1}\psi_{3} &= E_{\rm OBC}\psi_{2},\\
			t_{-2}\psi_{L-2} + t_{-1}\psi_{L-1}&= E_{\rm OBC}\psi_{L}.
			\label{bc}
		\end{split}
	\end{equation}
	Then, by substituting the general solution~\cite{nhse5}:
	\begin{equation}
		\psi_{n}= \sum_{j=1}^{3} c_j \beta_j^n,
	\end{equation}
	into Eq.~(\ref{bc}), one can reduce the problem into the following matrix equation:
	\begin{equation}
		H_B \left( \begin{array}{c}
			c_1 \\
			c_2 \\
			c_3
		\end{array}
		\right)=\left(\begin{array}{cccc}
			A_1\beta_1 & A_2\beta_2 & A_3\beta_3\\
			B_1\beta_1^2 & B_2\beta_2^2  & B_3\beta_3^2  \\
			C_1\beta_1^L  & C_2\beta_1^L & C_3\beta_1^L 
		\end{array} \right)\left( \begin{array}{c}
			c_1 \\
			c_2 \\
			c_3
		\end{array}
		\right)=0, 
		\label{mat}
	\end{equation}
	where
	\begin{equation}
		\begin{split}
			A_j &=t_1 \beta_j -E_{\rm OBC},\\
			B_j&= t_{-1}/\beta_j - E_{\rm OBC} + t_1\beta_j,\\
			C_j&= t_{-2}/\beta_j^{2} t_{-1}/\beta_j - E_{\rm OBC}.
		\end{split}
	\end{equation}
	Then the condition for Eq.~(\ref{mat}) to have nontrival solutions is written as~\cite{band}
	\begin{equation}
		\text{det} [H_B]=\sum_{i\ne j\ne k=1}^{3}\epsilon_ {ijk}A_iB_jC_k\beta_i\beta_j^2\beta_k^L=0,
		\label{det}
	\end{equation}
	where the leading term can be ordered by
	\begin{equation}
		\beta_3^L\ge \beta_2^L\ge \beta_1^L,
	\end{equation}
	in the thermodynamic limit since other terms are independent of $L$. Here, $E_{\rm OBC}$ will form a continuous band in the thermodynamic limit if Eq.~(\ref{det}) has two leading terms (i.e., $\beta_3^L=\beta_2^L$); otherwise, there only exist finite solutions of $E_{\rm OBC}$. Thus, the GBZ of the model in Eq.~(\ref{eq2}) is determined by $|\beta_2|=|\beta_3|$. Finally, we can solve Eq.~(\ref{eq3}) numerically and then pick the solutions that satisfy the condition $|\beta_2|=|\beta_3|$, which form the GBZ in the complex plane.

	\begin{figure}[b]
		\centering
		\includegraphics[width=0.48\textwidth]{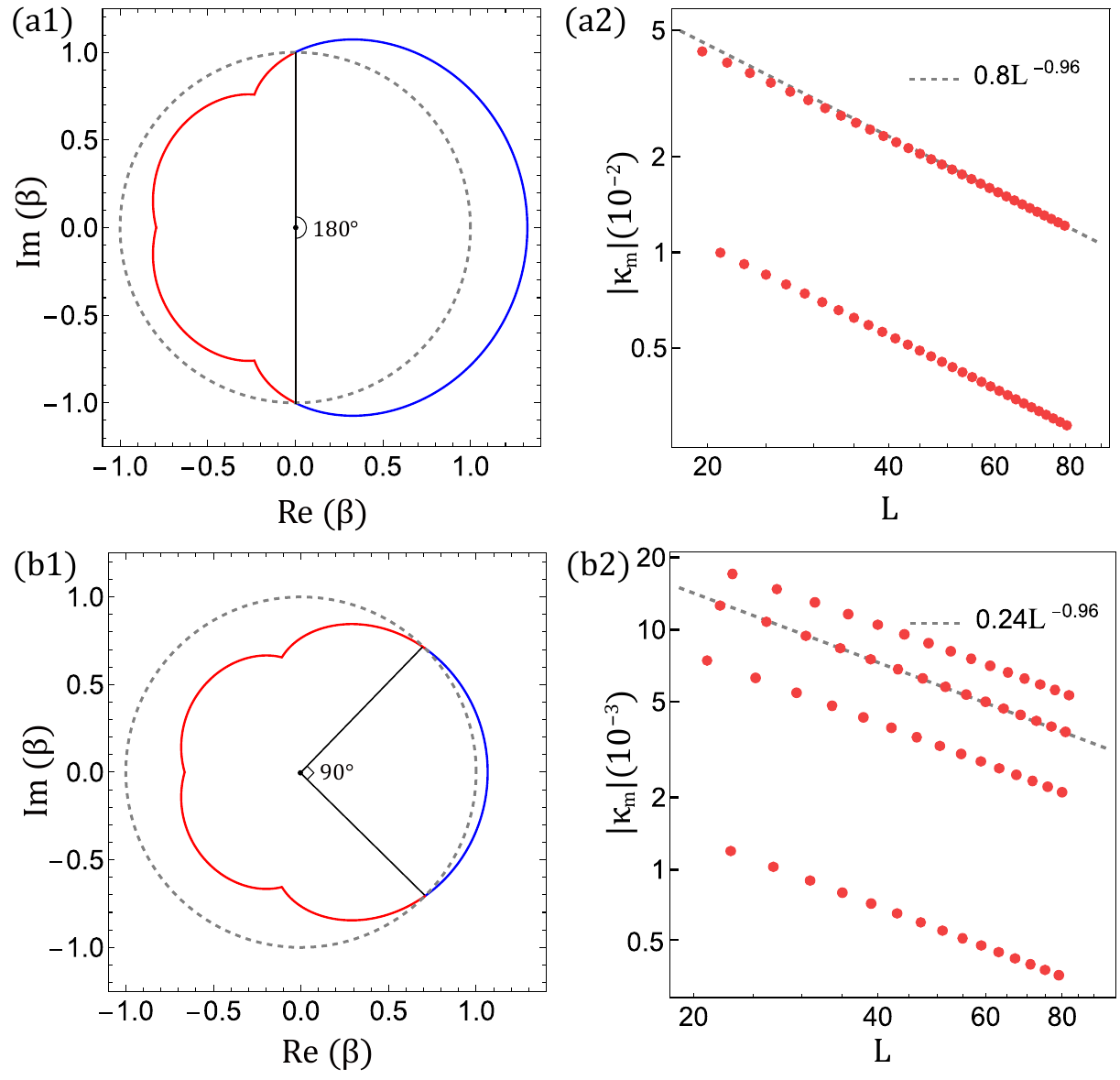}
		\renewcommand{\thefigure}{\arabic{figure}}	
		\caption{Generalized Brillouin zone (GBZ) and the scalefree power-law behaviors of inverse localization length near the Bloch point are plotted for (a1) and (a2) $H\left(\beta\right)=\beta^{-2}/2 +\beta^{-1}+\beta$, and (b1) and (b2) $H\left(\beta\right)=\beta^{-2}/2 +\beta^{-1}+(2+\sqrt{2})\beta/2$. The argument difference between the two intersections of GBZ and Brillouin zone (BZ) determines the periodicity of $\kappa_m$ with respect to system size $L$.}
		\label{FigS2}	
	\end{figure}   
	
	\section{Periodicity of the variation of $\kappa_m$ with $L$}
	In this appendix, we elucidate the periodic behavior of $\kappa_m$ with respect to changes in $L$. Generally, as $L$ increases, $\kappa_m$ exhibits periodic oscillations rather than following a monotonic trend. The complete periodicity of $\kappa_m$ is clearly observed when $L$ increases by $1$ unit each time. Figures~\ref{FigS2}(a2) and {FigS2}(b2) display that, as $L$ increases, $\kappa_m$ decreases while exhibiting significant periodic fluctuations. However, the main text plots, such as Fig.~\ref{fig2}(b), do not depict the periodic variation of $\kappa_m$ with $L$ because the interval of $L$ was set to a specific value.
	
	Additionally, we have ascertained that the periodicity of $\kappa_m$ as it varies with $L$ is determined by the intersection points of the GBZ and the BZ. Specifically, the period is calculated as $2\pi/\Delta\theta$, where $\Delta\theta$ ($\leq\pi$) represents the difference between the arguments of the two intersection points of the GBZ and the BZ. As shown in Fig.~\ref{FigS2}, a $\Delta\theta$ value of $\pi$ ($\pi/2$) results in a period of $2$ ($4$). There are certain special cases to consider: if $2\pi$ is not an integer multiple of $\Delta\theta$, the variation of $\kappa_m$ with $L$ may not present a clearly defined periodic pattern but still displays scalefree power-law behavior; if the GBZ intersects with the BZ only once, the period will be $1$. Furthermore, the periodicity of $|E_B-E_m|$ as it varies with $L$ follows a similar pattern.
	
	\begin{figure}[t]
		\centering
		\includegraphics[width=0.45\textwidth]{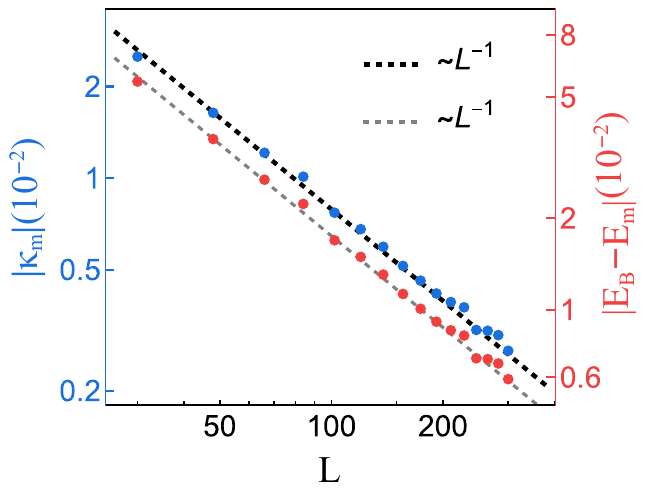}
		\renewcommand{\thefigure}{\arabic{figure}}	
		\caption{Scalefree power law behaviors with weak disorders. The model is pictorially shown in Fig.~\ref{fig4}(a), with parameters $t1=t2=1$, $t3=\frac{1}{5}$, $\gamma=\frac{1}{4}$. However, the intracell hopping coefficients within each unit cell is randomly generated from a uniform distribution of $[t_1-\delta t, t_1+\delta t]$, where $\delta t=\frac{1}{100}$.}
		\label{FigS3}	
	\end{figure}
	
	\section{Stability of the scalefree NHSE against disorder}
	
	In this appendix, we examine the emergence of scalefree power-law behaviors in the presence of weak disorders. We employ the same model and parameters used in Figs.~\ref{fig4}(b1)--\ref{fig4}(b3), with the exception that the intracell hopping coefficients within each unit cell are randomly generated from a uniform distribution $[t_1-\delta t, t_1+\delta t]$, where $\delta t$ represents the amplitude of the disorder~\cite{TI,sun}. Due to the presence of the disorder, the Bloch point and the energy spectrum in the thermodynamic limit are not accessible. Therefore, we choose the Bloch point from the clean system. In Fig.~\ref{FigS3}, it is evident that the scalefree power-law behaviors are still present even for $\delta t=\frac{1}{100}$.

\end{document}